\documentclass[11pt]{article} 
\usepackage[utf8]{inputenc}  
\usepackage{libertine}
\usepackage[left=2cm,right=2cm,top=2cm,bottom=2cm]{geometry}
\usepackage[pdftex]{graphicx} 
\usepackage{amssymb} 
\usepackage{amsmath} 
\usepackage{amsthm}
\usepackage{wrapfig} 
\usepackage{calc} 
\usepackage{footmisc} 
\usepackage[table, svgnames, dvipsnames]{xcolor}
\usepackage{makecell, cellspace, caption}
\usepackage{subcaption}
\usepackage{mathtools}
\usepackage{url}
\usepackage{hyperref}
\hypersetup{colorlinks,linkcolor={},citecolor={blue},urlcolor={red}}

\usepackage{lmodern}

\usepackage{booktabs}
\usepackage[square,sort,comma,numbers]{natbib}

\usepackage{epigraph}
\usepackage[T1]{fontenc}

\usepackage{etoolbox}
\usepackage{authblk}

\makeatletter

\patchcmd{\maketitle}{\@fnsymbol}{\@alph}{}{}
\makeatother

\title{A phenomenological estimate of the Covid-19 true scale \\ from primary data}

\author[a]{Luigi Palatella}
\author[b,c,d]{Fabio Vanni}
\author[d,e]{David Lambert} 

\affil[a]{Liceo Scientifico Statale ``C. De Giorgi'', Lecce, Italy}
\affil[b]{Sciences Po, OFCE , France}
\affil[c]{Institute of Economics, Sant'Anna School of Advanced Studies, Pisa, Italy}
\affil[d]{Department of Physics, University of North Texas, USA }
\affil[e]{Department of Mathematics, University of North Texas, USA }

\date{}
\begin{document} 
	\maketitle 
	
	\begin{abstract} 
		Estimation of the prevalence of undocumented SARS-CoV-2 infections is critical for understanding the overall impact of CoViD-19, and for implementing effective public policy intervention strategies. We discuss a simple yet effective approach to estimate the true number of people infected by SARS-CoV-2, using raw epidemiological data reported by official health institutions in the largest EU countries and the USA. 
	\end{abstract} 
	
	
	\section{Introduction} 
	As the coronavirus disease 2019 (CoViD-19) epidemic reached every corner of the world, each country has adopted different interventions to manage the early and long-term phases of the spread of the epidemic. This epidemic has forced many countries to react by imposing policies aimed at reducing population mobility together with internal and international border limitations. These policies have been informed by various measures of epidemic risk that all trace back to the number of new cases that the country's healthcare system has found each day.   Unrecorded and unnoticed cases make it difficult to estimate the true number of infected persons present at a given time (prevalence).

	Cases of infection are usually detected through testing, typically occurring when ill people (or their recent contacts) seek healthcare. Official data are mainly collected with medical swabs. This favors the examination of patients showing clear symptoms. However, key features of CoViD-19's dynamics have to do with asymptomatic or pre-symptomatic transmission \cite{meyerowitz2020towards, moghadas2020implications}. Transmission of severe acute respiratory syndrome coronavirus 2 (SARS-CoV-2) can occur before symptom onset in the infector, which presents a stumbling block to efforts to stop the spread of the disease. Infected persons often do not develop noticeable symptoms until after the viral latent period, the time between being infected with the virus and first becoming contagious. (This differs from the incubation period: the time between being infected with the virus and first developing symptoms.) Recent studies \cite{peirlinck2020outbreak, lauer2020incubation, Byrne2020} have found that SARS-CoV-2 has the notable property that the latent period of SARS-CoV-2, is shorter than the incubation period of the virus. Thus, individuals can become contagious even before they show symptoms. Early in the epidemic, and in various situations afterward, diagnosis has been based on having a certain set of symptoms.  More commonly, an infection diagnosis comes from direct detection of SARS-CoV-2 RNA by nucleic acid amplification tests, typically reverse-transcription polymerase chain reaction (RT-PCR) from the upper respiratory tract. Furthermore, testing capacity depends on the demands a healthcare system is under. So that the confirmed case numbers reported during an outbreak represent only a fraction of the true levels of infection in a community.

	Undocumented infections often are not detected due to mildness, limitedness, or absence of symptoms. As they are not quarantined, they can expose far more of the population to the virus, thereby sustaining the spread of the epidemic.  Moreover, the fraction of undocumented (but still infectious) cases is a critical epidemiological feature that modulates the pandemic potential of a virus.  The ability to estimate the scale of an epidemic is of paramount importance medically, socially, and economically, as it affects viral hot-spot detection, resource allocation, and intervention planning. Restrictive measures have indeed exempted firms producing essential goods and services. In addition, companies able to massively employ remote work have succeeded in mitigating the negative effects. Better estimation of the true prevalence of CoViD-19 cases is crucially important to set the strength and scale of non-medical interventions and improve the evaluation of the economic and health impacts associated with lockdown and reopening policies.

	Many modeling methods use the mortality rate to estimate the scale of CoViD-19 \cite{flaxman2020estimating, li2020substantial, oke2020global, phipps2020robust}. These methods rely on fixed estimates of certain epidemiological properties (such as the onset-to-death interval distribution and the generation-time distribution). Further, they assume that the fatality rate is the same everywhere and that its true value is near the lowest observed value ($\simeq 0.5-1.0\%$). Finally, these methods require a long time to produce results.  
	We propose an independent way to estimate the true incidence of SARS-CoV-2. The intuition behind this approach lies in the well-known effect that depletion of the susceptible population reduces the reproduction number $R_t$. 
	We make a quantitative analysis of this saturation effect to estimate the true scale of the Covid-19 pandemic, i.e., the order of magnitude of the actual number of people that have been infected.

	In the Methods section, we explain the mathematical background for the estimation of the fraction of un-diagnosed cases by using both the number of new positive tests and the instantaneous reproduction number $R_t$. In the Data and Results section, we show the results for Italy, France, Spain, Germany, and USA. In the Conclusions section, we summarized our research and conclude the paper.
	
	\section{Methods} 
	At the very beginning of an epidemic, the fraction of the population that is susceptible to the disease is 1. The basic reproduction number, $R_0$, is the average number of people that are infected by a single infected person per day when everyone is susceptible.  In general, $R_0$ depends on the capacity of the virus to infect people, as well as on aspects that vary from country to country such as social habits, movement patterns, average health status, and sanitation. Moreover, the reproduction number may depend on time due to seasonal conditions, air humidity, UV exposure, and social changes (e.g., an increase in mask wearing or increased average interpersonal distance). When people become immune or die the susceptible fraction of the population decreases. All else being equal, this leads to a decrease in the instantaneous (or apparent, or actual) reproduction number, $R_t$. 
	This phenomenon is described by the following relation: 
	\begin{equation}\label{eq_1}
	R_t  = R_0(t) \left(1 - C(t) \right). 
	\end{equation} 
	Here, $C(t)$ is the true {\em fraction} of the population that has been infected at time $t$. Let us suppose that the testing system of a region is not able to detect every positive case, we define 
	\begin{equation} 
	\lambda(t) \equiv \frac{\mathrm{detected\: cases}}{\mathrm{actual \: cases}} 
	= \frac{c(t)}{C(t)},
	\end{equation} 
	where $c(t)$ is the total number of detected cases (the cumulative number of cases) as a fraction of the whole population.

	The instantaneous reproduction number, $R_t$, is an important tool for detecting changes in disease transmission over time \cite{fraser2007estimating, gostic2020practical}. Policy makers and public health officials use $R_t$ to assess the effectiveness of interventions and to inform policy. Specifically, the instantaneous reproduction number measures disease transmission at a given point in time, $t$.  One can interpret it as the average number of people that each contagious individual at time $t$ would infect per day, if conditions remained unchanged. We assume, following \cite{nishiura2010time, champredon2018equivalence, breda2012formulation}, that $R_t$ depends on the number of new cases per day, $j(t)$, only through $1-C(t)$ (as in Eq.\eqref{eq_1}) and that $j(t)$ obeys the so-called renewal equation: 
	\begin{equation} 
	j(t)=\frac{d}{dt} C(t)=\int_{0}^{\infty}A(t, \tau) j(t-\tau) d\tau +i(t),
	\end{equation}
	where $i(t)$ is the number of imported cases per day, and $\tau$ represents the infection age, the amount of time since an individual became infected.

	The function $A(t, \tau) $ is a product of $R_t$ and the generation-time distribution $g(t,\tau) $\cite{cori2013new}. The distribution $g(t,\tau)$ represents the probability density for a person infected at time $t-\tau$ to be infectious at time $t$ (as with any probability density $\int_0^\infty g(t,\tau)d\tau=1$). For simplicity, we assume that this distribution is independent of $t$, i.e., $g(t,\tau)=g(\tau)$, so that  
	\begin{equation} 
	j(t)=R_t\int_{0}^{\infty}g (\tau) j(t-\tau) d\tau. 
	\end{equation} 
	Typically, the generation distribution is unknown, though it can be approximated by assuming it is the same as the \textit{serial-interval distribution,} which refers to the time between successive cases in a chain of transmission. We follow \cite{flaxman2020report} in using a gamma function given by 
	\begin{equation} 
	g(\tau) = \beta^{\alpha} \tau^{\alpha-1} \exp (-\beta \tau)  
	\end{equation} 
	with $\alpha = 1.87$ and $\beta = 0.28$.  We estimate $R_t$ by \cite{fraser2007estimating} 
	\begin{equation}\label{Rteq} 
	R_t = \frac{j(t)}{\int\limits_{0}^{\infty}g (\tau) j(t-\tau) d\tau}\approx\frac{\tilde{j}(t)}{\int\limits_{0}^{\infty}g (\tau) \tilde{j}(t-\tau) d\tau},
	\end{equation} 
	where $\tilde{j}(t)\equiv\frac{dc}{dt}=\frac{d\lambda(t)}{dt}C(t)+\lambda(t)j(t)$.  We assume that $\lambda(t)$ changes slowly enough that $\tilde{j}(t)\approx\lambda(t)j(t)$ and $\frac{\lambda(t-\tau)}{\lambda(t)}\approx 1$ (for $0\leq\tau\lesssim\frac{1}{\beta}$), so that the approximation in Eq.\eqref{Rteq} holds.

	Starting from the saturation effect of Eq.\eqref{eq_1}, we wish to find two unknown variables namely, $R_0$ and $\lambda$. With ideal data, one could immediately apply linear fitting techniques. However, the data reported by most countries are affected by a strong weekly fluctuation. To reduce this effect, we perform a moving average of the $R_t$ obtained over 7 or 14 days (the results are almost identical). 

	
	As Fig.\ref{fig_L1} shows, $R_0(t)$ abruptly increased at the end of September, then decreased in the first days of November. Since $R_0(t)$ varied significantly, a simple linear fit with $R_0$ as a parameter would yield poor results. Instead we make an independent estimate of $R_0(t)$ by doing a weighted average of the $R_t$'s obtained in all the regions in a country at a given calendar time $t$. This estimate reads as: 
	\begin{equation}\label{eq_fluct} 
	\langle{R_0}(t) \rangle = \frac{1}{N_\mathrm{regions}} \sum_{\mathrm{k \in regions}} R^{(k)}_t 
	\frac{1}{1 - c^{(k)}(t)}, 
	\end{equation}  
	where $c^{(k)}(t)$ is the fraction of the population recorded as having been infected in region $k$ up to time $t$, while $R^{(k)}_t$ is the value obtained using Eq.(\ref{Rteq}) at time $t$ in region $k$\footnote{One should use instead $\langle{R_0}(t) \rangle = \frac{1}{N_\mathrm{regions}} \sum_{\mathrm{k \in regions}} R^{(k)}_t\frac{1}{1 - c^{(k)(t)}/\lambda} $ but the value of $\lambda$ at this stage is unknown. We use a recursive approach inserting the value obtained in the linear fit procedure explained below and then repeating the whole procedure. In any case we see a negligible difference in the value of $\lambda$ obtained by using the simplest factor $1 - c^{(k)} (t)$ as in Eq.(\ref{eq_fluct}), so we present the simplest approach.}.  We apply a linear fit to $R_t / \langle{R_0}(t) \rangle$ as a function of $c(t)$. Inserting Eq.\eqref{eq_fluct} into Eq.\eqref{eq_1}, we obtain
	
	\begin{equation}\label{fit2} 
	\frac{R_t}{ \langle{R_0}(t) \rangle}  = \left (1 - C(t) \right) = \left (1 - \frac{c(t)}{\lambda} \right).
	\end{equation}
	So we interpret the intercept of the fit as an extrapolation of the initial susceptible population fraction and the slope as $\frac{1}{\lambda}$. 
	
	\section{Data and Results}  
	In this section, we apply our method to data and obtain satisfactory results. Our data sources are listed in Table \ref{tabella} below.
	\begin{table}[!ht] 
		\centering 
		\begin{tabular}{  r | l } 
			\textit{country} &\textit{ data source} \\ 
			\bottomrule 
			Italy & Dipartimento della Protezione Civile \cite{ProtezCivileCov} and Istituto Superiore di Sanit\`a \cite{epicentro}\\ 
			France & Sant\'e publique France \cite{francedata}\\ 
			Spain & Ministerio de Sanidad \cite{spaindata} and Datadista \cite{spaindata2} \\ 
			Germany & Robert Koch Institute \cite{germandata1} and  Risklayer GmbH \cite{germandata2} \\ 
			UK & The official UK Government website \cite{UKgov}\\ 
			USA & The Covid Tracking Project \cite{CovidTracking}\\ 
			\bottomrule 
		\end{tabular} 
		\caption{\label{tabella} Summary of primary data sources available both at national and "regional" level.} 
	\end{table} 
	The analysis was performed at the regional level in all the countries, specifically: {\em Regione} for Italy, {\em Region} for France, {\em L\"ander} for Germany, {\em Comunidad aut\'onoma} for Spain, {\em Federal States} for the USA, and {\em Regions of England plus Wales, Scotland, and Northern Ireland} for the UK. For each region, the total population used to calculate $c(t)$ was retrieved from the most recent census available online (except for regions of France which report population size directly in the dataset).  All the data files offer extended epidemiological reports; however, we only used the number new CoViD-19 cases per day.

\begin{figure} 
		\includegraphics[width = \textwidth]{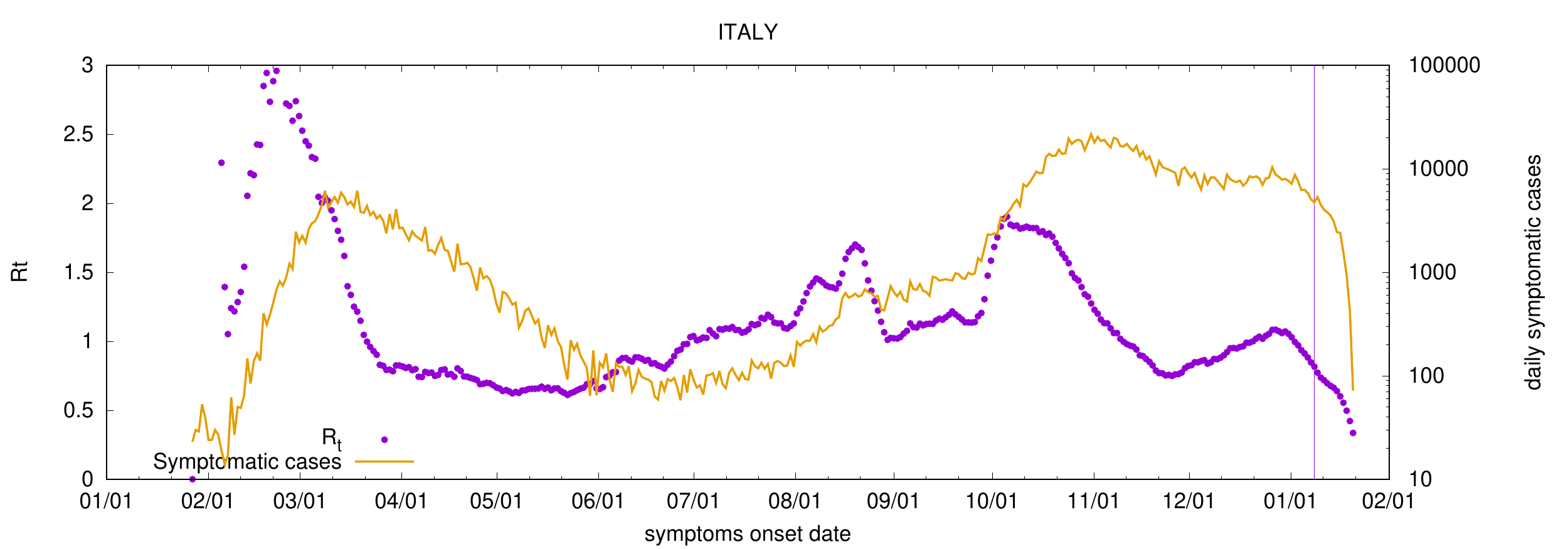} 
		\caption{Number of symptomatic cases in Italy per day  
			reported by date of onset of symptoms (right axis, log scale) and $R_t$ (left axis, linear scale) 
			obtained using Eq.(\ref{Rteq}) and then averaged over a week. The vertical line indicates
			the most recent data reported as consolidated.}
		\label{fig_onset} 
	\end{figure}

	In some cases, official institutions supply data by date of onset of symptoms, as in the case of Italy, see Fig.\ref{fig_onset}.
	This type of data is especially nice for finding the instantaneous reproduction number $R_t$. Indeed, Italian {\em Istituto Superiore di Sanit\`a} currently uses these values because they are virtually unaffected by fluctuations in testing capacity. Unfortunately, these data are only available for the whole country rather than particular regions. So, we only used these data to compare our measure of $R_t$ with the one based on the raw data of new infections. We find that our measure is accurate up to reasonable statistical fluctuations.

\begin{figure}[!ht] 
		\includegraphics[width = \textwidth]{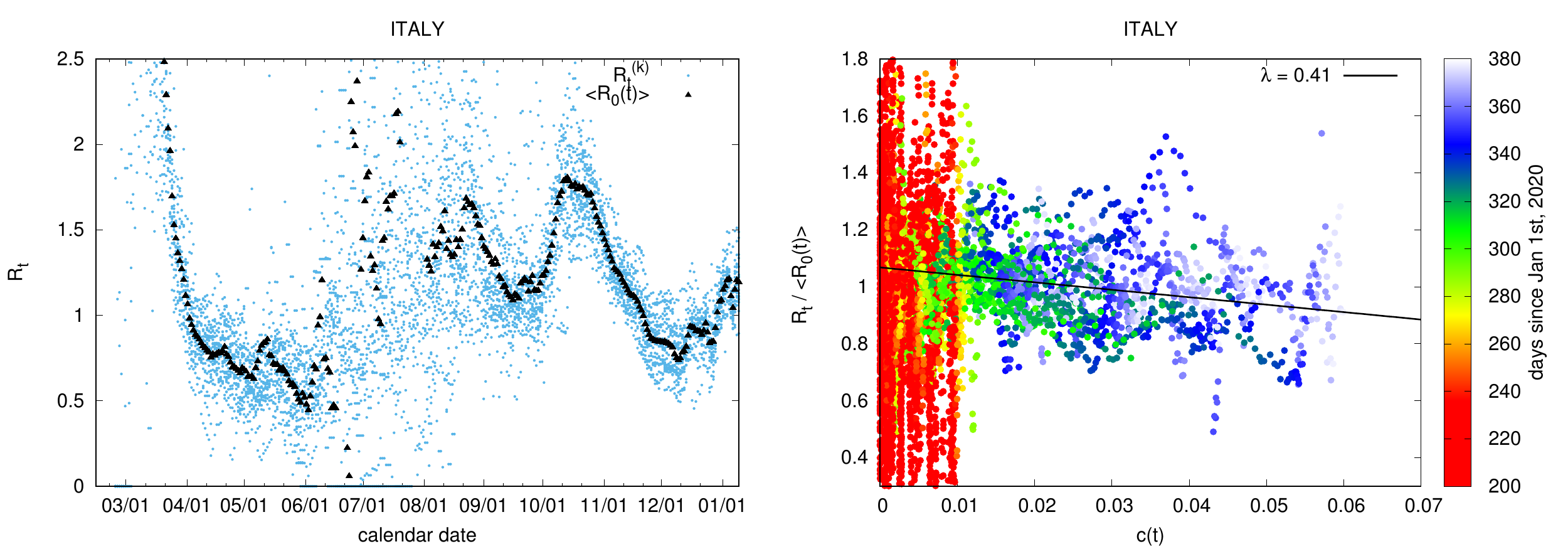} 
		\includegraphics[width = \textwidth]{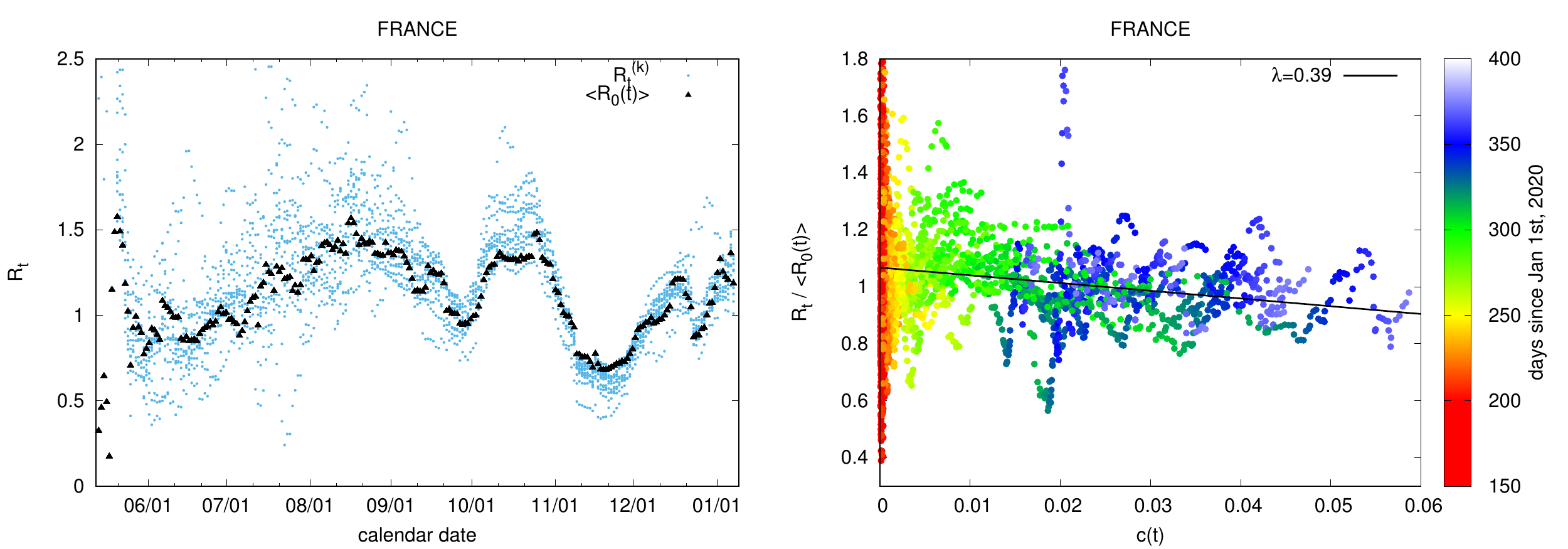} 
		\includegraphics[width = \textwidth]{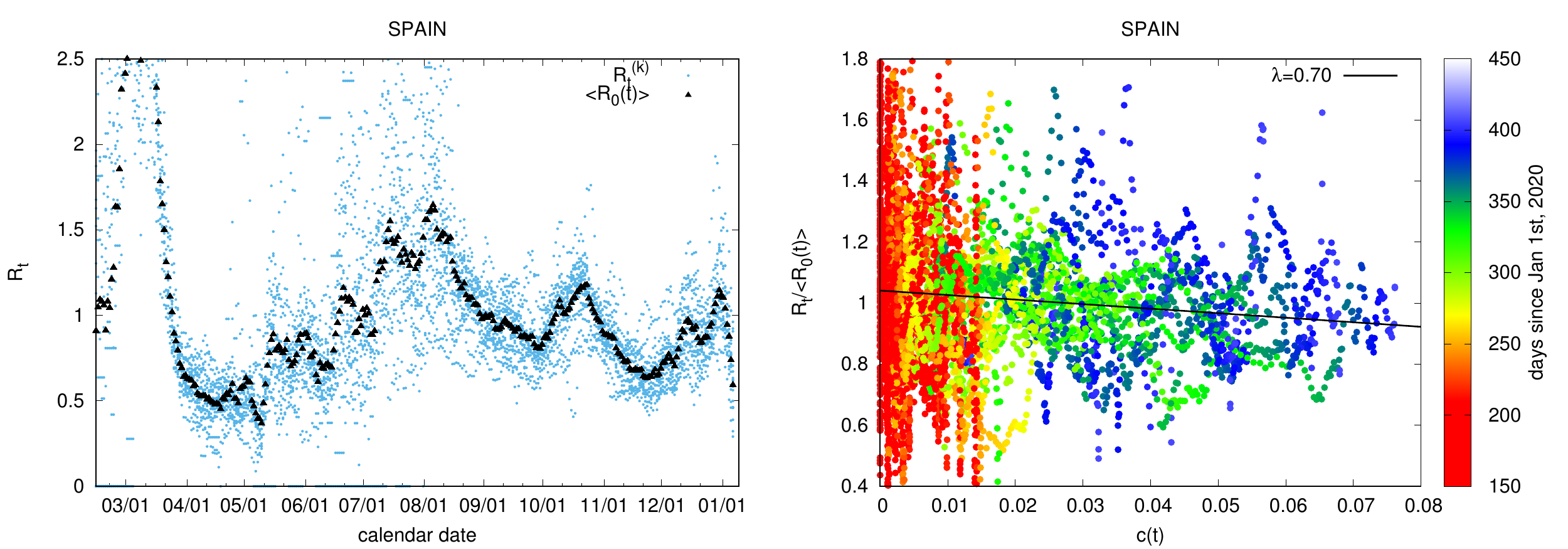} 
		\caption{(Color online) estimation results for Italy, France, and Spain. (Left panel) scatterplot of $R_t$ evaluated for all 
			the regions together with $\langle R_0(t)\rangle$. For the Italian data notice that, except for summer, $R_t$ evaluated on standard data is quite like that evaluated on the ISS data reporting cases by date of onset of symptoms. 
			(Right panel) The fluctuations of $R_t^{(k)}/\langle R_0(t)\rangle$ vs $c(t)$ with the best linear fit leading the estimate of $\lambda$. Color indicates date in days from Jan. $1^\text{st}$, 2020.} 
		\label{fig_L1} 
	\end{figure}

\begin{figure}[!ht] 
		\includegraphics[width = \textwidth]{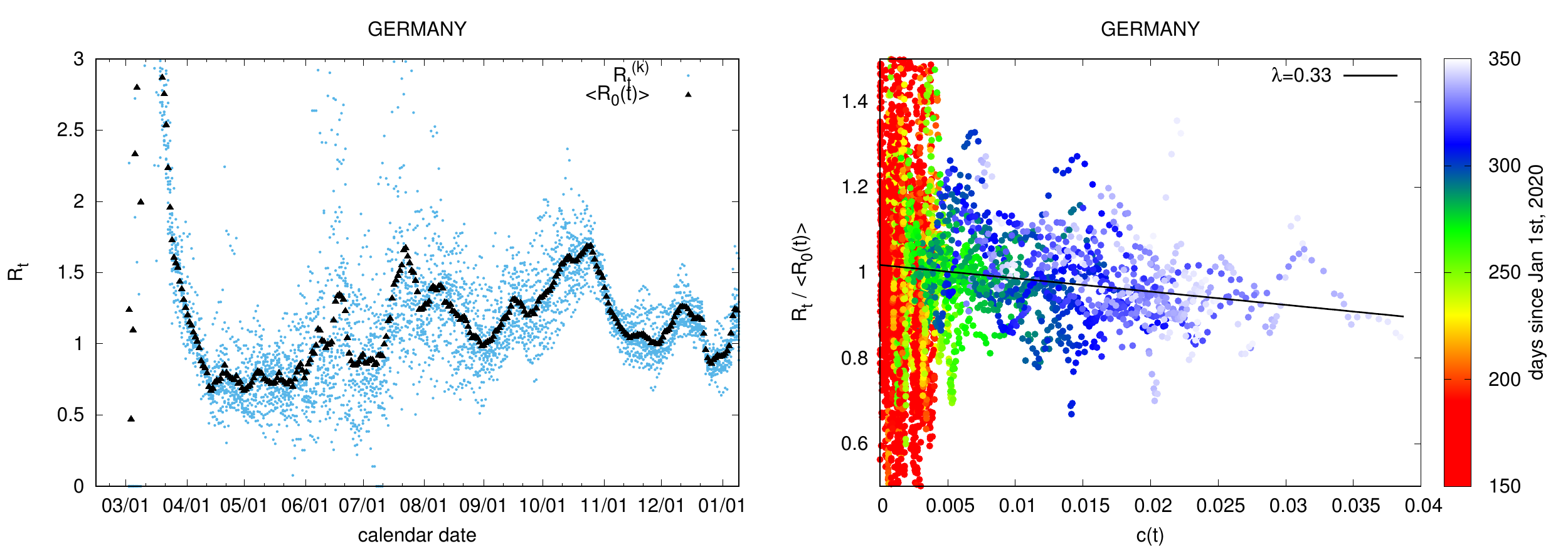} 
		\includegraphics[width = \textwidth]{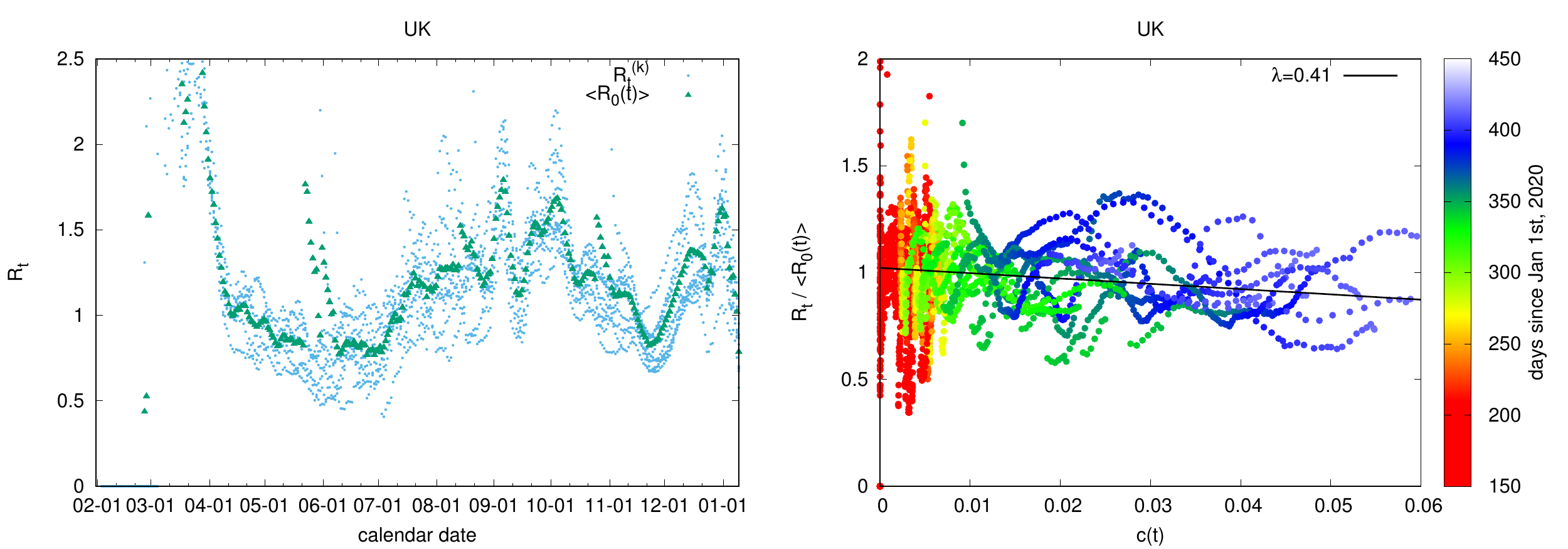} 
		\includegraphics[width = \textwidth]{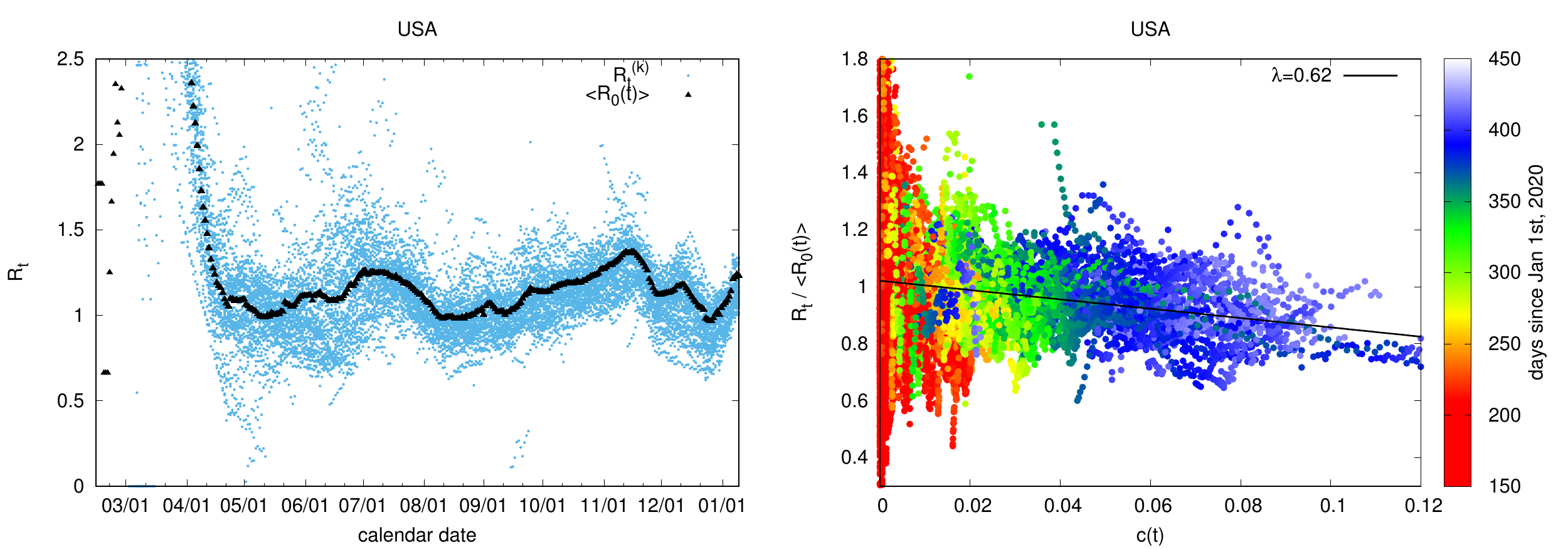} 
		\caption{(Color online) estimation results for Germany, UK, and USA. (Left panel) scatterplot of $R_t$ evaluated for all the regions together with $\langle R_0(t)\rangle$.
			(Right panel) The fluctuations of $R_t^{(k)}/\langle R_0(t)\rangle$ vs $c(t)$ with the best linear fit leading the estimate of $\lambda$. The color shows the date as days from Jan. $1^\text{st}$, 2020.} 
		\label{fig_L2} 
	\end{figure}

	Now we turn to Fig.\ref{fig_L1} and Fig.\ref{fig_L2}. For each country we plot the value of $R^{(k)}_t$ for its regions versus time and $R_t / \langle{R_0}(t) \rangle$ versus $c(t)$ along with the best fit for $\lambda$. The fit was performed neglecting the first $100$ days of the epidemic when testing was quite irregular and the number of PCR tests extremely low. The error on $\lambda$ was evaluated by changing this threshold from $50$ to $150$ days. 
	\clearpage 
	A straightforward application of the estimate of the true number of infections is to adjust the case fatality rate (CFR) to better approximate the infectious fatality rate (IFR).  The CFR is defined as the number of reported deaths per number of reported cases. It approximates IFR, which is an estimate of the death rate among all those infected with SARS-CoV-2, the virus that causes CoViD-19. In table \ref{cfr} we estimate the unrecorded fraction of the true number of infected individuals as well as the corrected value of the case fatality rate ($\lambda$-CFR).
	This correction does not capture every reason the CFR can differ from the IFR. Some additional differences are due to delayed effects, population age, typical medical conditions of the population, and healthcare system efficiency. 
	\begin{table}[!ht]
		\begin{center}
			\begin{tabular}{ c   c  c  || c  c }
				\textit{Country} & $\lambda$ & C.I.& CFR (\%) & $\lambda$-CFR (\%)\\
				\midrule
				Italy &  $0.35$ & {\small $[0.31 , 0.42]$} & $3.52$ & $1.23$ \\
				France  & $0.39$ &{\small $[0.34, 0.46]$} & $2.45$ &  $0.95$\\
				Germany & $0.33$  &{\small $[0.31, 0.41]$} & $2.10$ & $0.69$ \\
				Spain & $0.70$ &{\small  $[0.63, 0.84]$ }& $2.74$ & $1.92$\\
				UK & $0.41$ &{\small $[0.37 , 0.49]$ }& $2.70$ & $1.11$ \\
				USA & $0.62$ &{\small $[0.56,0.74]$ }& $1.70$& $1.05$ \\
				\bottomrule
			\end{tabular}
		\end{center}
		\caption{ Results of the testing variable estimation as for January 2021.}
		\label{cfr}
	\end{table}

	As the vaccination campaign continues, we must consider the effects of immunization through vaccination on our analysis. The functional form will change to $R_t/\langle R_0\rangle \sim 
	\left( 1-{c(t)}/{\lambda} \right ) \left( 1-\nu(t) \right) \simeq
	\left(1-{c(t)}/{\lambda}-\nu(t)\right) $, where new term takes in account the vaccinated fraction of the population $\nu(t)$.  If data on the vaccination campaign are available at the regional level, and $\nu(t)$ reaches a substantial value, one should perform a two-dimensional fit of $R_t/\langle R_0\rangle$ as a function of $c(t)$ and $v(t)$.  The slope of the iso-$R_t/\langle R_0\rangle$ lines, i.e., the lines in the $c(t)-v(t)$ plane of constant $R_t/\langle R_0\rangle$, will give another estimate of $\lambda$. This will allow us to compare and contrast the effects of vaccination and new infectious cases on $R_t$. If $\lambda \simeq 1$, their effects should be the same. If $\lambda < 1$, we expect that the effect of a given number of infectious individuals on $R_t$ to be {\em larger} than that of the same number of vaccinated. 
	
	\paragraph{Robustness check using seroprevalence data} 
	Our findings can be checked by comparing our results with alternative methods used to evaluate the true number of people that have had SARS-CoV-2. One of the most reliable methods is the analysis of seroprevalence of IgG antibodies in blood. We compare our $\lambda$ values with those obtained by two papers \cite{havers2020seroprevalence, bajema2020comparison} and a dashboard on an official website \cite{dashboard}.

\begin{figure} 
		\includegraphics[width = 0.5\textwidth]{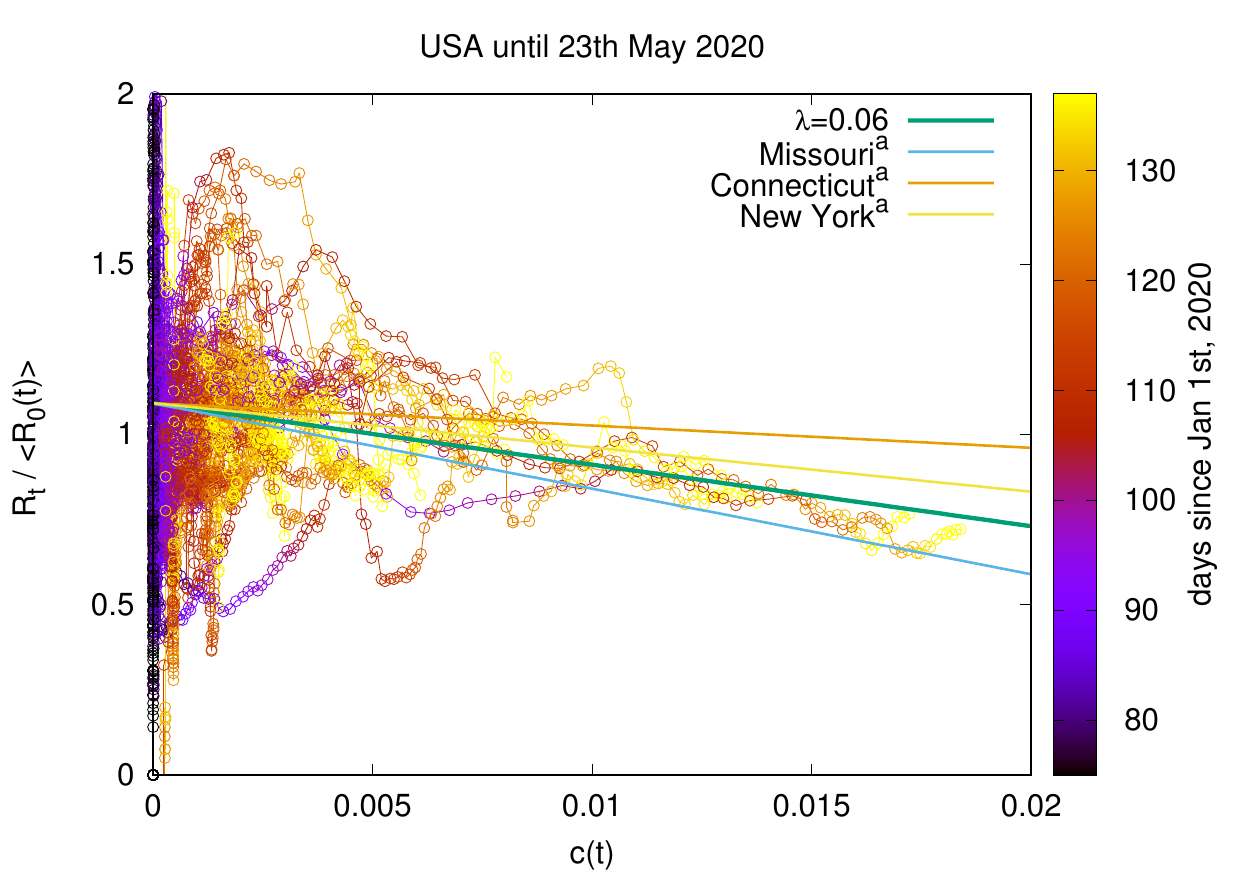} 
		\includegraphics[width = 0.5\textwidth]{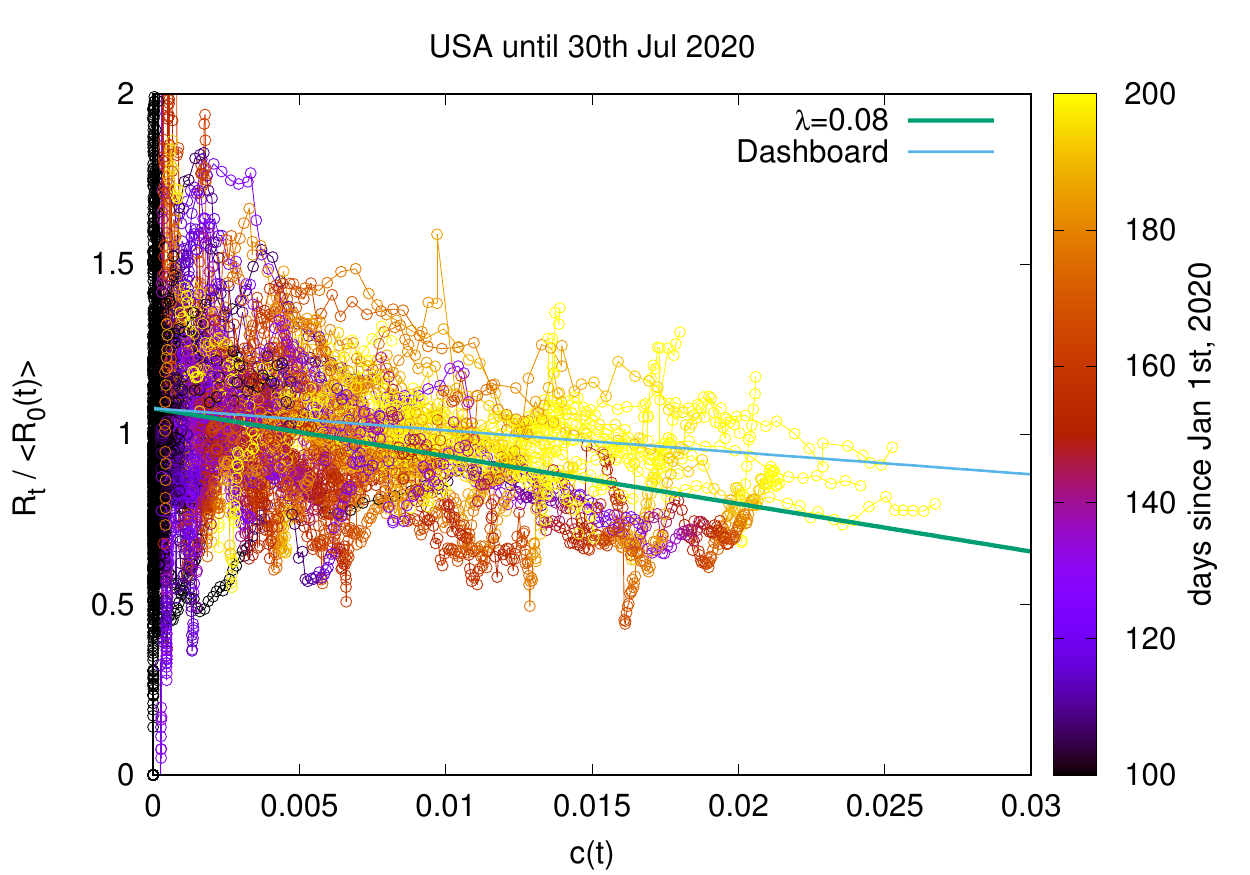} 
		\caption{\label{sero} 
			(left panel, color online) best fit value for $\lambda$ through May $23^\text{rd}$, 2020 compared with $\lambda^a$ obtained from Table 3 of \cite{havers2020seroprevalence} for New York ($\lambda = 11.9$) and for the extremal cases of Missouri ($\lambda = 1/23.8$) and Connecticut ($\lambda = 1/6.0$). (Right panel, color online) same approach as in the left panel with data ending July $30^\text{th}$, 2020 compared with the average value given by \cite{dashboard} ($\lambda = 1/6$)} 
	\end{figure}

	In particular, \cite{havers2020seroprevalence} reports the fraction of people showing IgG antibodies to SARS-CoV-2 over the number of officially reported cases as of the end of May $2020$. This ratio should be compared with $\lambda^{-1} $. In Table 3 of \cite{havers2020seroprevalence} data for 10 sites are reported. They range from a minimum of $6.0$ for Connecticut to a maximum of $23.8$ for Missouri. The worst hit state, New York, has a value of $11.9$.  In Fig.\ref{sero} (left panel), we report the value of $\lambda$ using data covering the same period, namely ending May $23^\text{rd}$, $2020$. We report our best fit yielding $\lambda = 0.06$, together with the line corresponding to the $\lambda$ obtained from serological data described in \cite{havers2020seroprevalence}. The same approach was followed in Fig.\ref{sero} (right panel), where we report the typical value retrieved from \cite{dashboard} (between 6 and 7 times the official number of cases) with our fit at the date of the last update of the website (July $2020$).
	
	Passing from late spring to the end of summer, $\lambda$ increased considerably. This effect is apparent in the dashboard data \cite{dashboard} and the same pattern holds in all countries analyzed. We presume this increase is due to increased testing ability and contact tracing efforts made in all countries. Nevertheless, as a final remark, we stress that if our results are confirmed, in all countries the tracing capacity is not enough, by itself, to mitigate the spread of SARS-CoV-2. Due to the large contribution of asymptomatic or mild-symptomatic cases, we think that reaching a value of $\lambda \simeq 1$ would be necessary. This would require a massive use of tests as seen, for example, in China during its prevention of a second wave.

	Finally, in comparing our approach to the seroprevalence data, we found that our method gives results that are of the same order of magnitude. Our method tends to overestimate the serological results. This could be mark against our approach, but one should note that seroprevalence-based infection estimates are conservative \cite{dashboard}. Some studies have found that infected persons who are asymptomatic or have mild symptoms do not have detectable antibodies. Other studies found that antibody levels decrease and become undetectable in some patients over time (see, for example, \cite{le2020sars, ng2020pre}). Another explanation of this overestimation is that the number of susceptible individuals is not the entire population due to cross-immunity with other common coronaviruses \cite{lipsitch2020cross, Ng2020, doshi2020covid}.
	
	\section{Conclusions} 
	Knowledge of true prevalence of CoViD-19 is critical for informing policy decisions about how to distribute resources and manage the impacts of CoViD-19 on public health, society and the economy\cite{organisation2020territorial, gans2020economic}. The true scale of the epidemic can affect economic development since it reduces long-run economic growth by limiting the size of social networks. 
	On the contrary a low prevalence estimate could lead people to take more epidemic-relevant risks, making disease eradication impossible using social distancing policies only.

	We have proposed a method to estimate the true number of infected people, unveiling the true scale of CoViD-19 by using PCR test data alone.
	We the used this method to estimate a more reliable case fatality rate.  Our approach is a phenomenological estimate of the true scale of the epidemic, since it is based on an empirical relationship between phenomena, in a way which is consistent with fundamental theory, but is not directly derived from that theory. Consequently, our method can be affected by errors which we are not able to distinguish from the lack of theoretical understanding. Nevertheless, there are two remarkable results. The few attempts to assesss the true impact of SARS-CoV-2 via serological tests yield $\lambda$-values comparable in order of magnitude with our calculations \cite{crocerossa}. Moreover, countries that performed better tracing, like the USA, 
	experienced a fatality rate lower than Italy, France, and Spain (in the first wave).  Improved contact tracing performance (higher $\lambda$) leads to a lower fatality rate because the rate is calculated using the correct denominator. Good contact tracing can also help keep the epidemic under control, preventing the virus from reaching fragile and elderly people. Thus leading to a bona fide decrease in the fatality rate.

	We have proposed that the study of the graph $R_t$ vs $c(t)$ and $R_t / \langle{R_0}(t) \rangle$  vs $c(t)$ can provide useful insight into epidemic dynamics. These methods should be useful in other epidemics with incomplete data due to the presence of unrecorded contagious individuals, for example, the seasonal flu. Minor modifications allow this approach to keep its effectiveness even in the presence of active vaccination campaigns. By contributing to a better-informed response to epidemics, we believe this work serves to benefit society as a whole.
	
	\section{Acknowledgments} 
	Fabio Vanni acknowledges support from the European Union's Horizon 2020 research and innovation programme under grant agreement No.822781 GROWINPRO - Growth Welfare Innovation Productivity. 
	
	\section{Conflict of interest} 
	The authors declare that there are no conflicts of interest regarding the publication of this paper. 
	
	\bibliographystyle{unsrt} 
	\bibliography{referX}  

\begin{thebibliography}{10}

\bibitem{meyerowitz2020towards}
Eric~A Meyerowitz, Aaron Richterman, Isaac~I Bogoch, Nicola Low, and Muge
  Cevik.
\newblock Towards an accurate and systematic characterisation of persistently
  asymptomatic infection with sars-cov-2.
\newblock {\em The Lancet Infectious Diseases}, 2020.

\bibitem{moghadas2020implications}
Seyed~M Moghadas, Meagan~C Fitzpatrick, Pratha Sah, Abhishek Pandey, Affan
  Shoukat, Burton~H Singer, and Alison~P Galvani.
\newblock The implications of silent transmission for the control of covid-19
  outbreaks.
\newblock {\em Proceedings of the National Academy of Sciences},
  117(30):17513--17515, 2020.

\bibitem{peirlinck2020outbreak}
Mathias Peirlinck, Kevin Linka, Francisco~Sahli Costabal, and Ellen Kuhl.
\newblock Outbreak dynamics of covid-19 in china and the united states.
\newblock {\em Biomechanics and modeling in mechanobiology}, page~1, 2020.

\bibitem{lauer2020incubation}
Stephen~A Lauer, Kyra~H Grantz, Qifang Bi, Forrest~K Jones, Qulu Zheng,
  Hannah~R Meredith, Andrew~S Azman, Nicholas~G Reich, and Justin Lessler.
\newblock The incubation period of coronavirus disease 2019 (covid-19) from
  publicly reported confirmed cases: estimation and application.
\newblock {\em Annals of internal medicine}, 172(9):577--582, 2020.

\bibitem{Byrne2020}
Andrew~William Byrne, David McEvoy, Aine~B Collins, Kevin Hunt, Miriam Casey,
  Ann Barber, Francis Butler, John Griffin, Elizabeth~A Lane, Conor McAloon,
  Kirsty O'Brien, Patrick Wall, Kieran~A Walsh, and Simon~J More.
\newblock Inferred duration of infectious period of sars-cov-2: rapid scoping
  review and analysis of available evidence for asymptomatic and symptomatic
  covid-19 cases.
\newblock {\em BMJ Open}, 10, 08 2020.

\bibitem{flaxman2020estimating}
Seth Flaxman, Swapnil Mishra, Axel Gandy, H~Juliette~T Unwin, Thomas~A Mellan,
  Helen Coupland, Charles Whittaker, Harrison Zhu, Tresnia Berah, Jeffrey~W
  Eaton, et~al.
\newblock Estimating the effects of non-pharmaceutical interventions on
  covid-19 in europe.
\newblock {\em Nature}, 584(7820):257--261, 2020.

\bibitem{li2020substantial}
Ruiyun Li, Sen Pei, Bin Chen, Yimeng Song, Tao Zhang, Wan Yang, and Jeffrey
  Shaman.
\newblock Substantial undocumented infection facilitates the rapid
  dissemination of novel coronavirus (sars-cov-2).
\newblock {\em Science}, 368(6490):489--493, 2020.

\bibitem{oke2020global}
Jason Oke and Carl Heneghan.
\newblock Global covid-19 case fatality rates. cebm.
\newblock 2020.
\newblock
  \url{https://www.cebm.net/covid-19/global-covid-19-case-fatality-rates/}.

\bibitem{phipps2020robust}
Steven~J Phipps, R~Quentin Grafton, and Tom Kompas.
\newblock Robust estimates of the true (population) infection rate for
  covid-19: a backcasting approach.
\newblock {\em Royal Society Open Science}, 7(11):200909, 2020.

\bibitem{fraser2007estimating}
Christophe Fraser.
\newblock Estimating individual and household reproduction numbers in an
  emerging epidemic.
\newblock {\em PloS one}, 2(8):e758, 2007.

\bibitem{gostic2020practical}
Katelyn~M Gostic, Lauren McGough, Edward~B Baskerville, Sam Abbott, Keya Joshi,
  Christine Tedijanto, Rebecca Kahn, Rene Niehus, James~A Hay, Pablo~M
  De~Salazar, et~al.
\newblock Practical considerations for measuring the effective reproductive
  number, r t.
\newblock {\em PLoS computational biology}, 16(12):e1008409, 2020.

\bibitem{nishiura2010time}
Hiroshi Nishiura.
\newblock Time variations in the generation time of an infectious disease:
  implications for sampling to appropriately quantify transmission potential.
\newblock {\em Mathematical Biosciences \& Engineering}, 7(4):851--869, 2010.

\bibitem{champredon2018equivalence}
David Champredon, Jonathan Dushoff, and David~JD Earn.
\newblock Equivalence of the erlang-distributed seir epidemic model and the
  renewal equation.
\newblock {\em SIAM Journal on Applied Mathematics}, 78(6):3258--3278, 2018.

\bibitem{breda2012formulation}
D~Breda, O~Diekmann, WF~De~Graaf, A~Pugliese, and R~Vermiglio.
\newblock On the formulation of epidemic models (an appraisal of kermack and
  mckendrick).
\newblock {\em Journal of biological dynamics}, 6(sup2):103--117, 2012.

\bibitem{cori2013new}
Anne Cori, Neil~M Ferguson, Christophe Fraser, and Simon Cauchemez.
\newblock A new framework and software to estimate time-varying reproduction
  numbers during epidemics.
\newblock {\em American journal of epidemiology}, 178(9):1505--1512, 2013.

\bibitem{flaxman2020report}
Seth Flaxman, Swapnil Mishra, Axel Gandy, H~Unwin, H~Coupland, T~Mellan, H~Zhu,
  T~Berah, J~Eaton, P~Perez~Guzman, et~al.
\newblock Report 13: Estimating the number of infections and the impact of
  non-pharmaceutical interventions on covid-19 in 11 european countries, 2020.

\bibitem{ProtezCivileCov}
DPC.
\newblock Covid-19 italia, 2020.
\newblock Sito del Dipartimento della Protezione Civile - Presidenza del
  Consiglio dei Ministri, \url{https://github.com/pcm-dpc/COVID-19}.

\bibitem{epicentro}
ISS.
\newblock Integrated surveillance of covid-19 in italy, 2020.
\newblock Scientific coordination by Centro Nazionale per la Prevenzione delle
  malattie e la Promozione della Salute, CNAPPS -
  ISS,\url{https://www.epicentro.iss.it/en/coronavirus}.

\bibitem{francedata}
Sant\'e publique France.
\newblock Taux d'incidence de l'\'epid\'emie de covid-19, 2020.
\newblock available at
  \url{https://www.data.gouv.fr/fr/organizations/sante-publique-france/}.

\bibitem{spaindata}
Ministerio de~Sanidad.
\newblock Enfermedad por nuevo coronavirus, covid-19, 2020.
\newblock available at
  \url{https://www.mscbs.gob.es/profesionales/saludPublica/ccayes/alertasActual/nCov/situacionActual.htm}.

\bibitem{spaindata2}
DATADISTA.
\newblock Coronavirus disease 2019 (covid-19) in spain, 2020.
\newblock available at
  \url{https://github.com/datadista/datasets/tree/master/COVID%2019}.

\bibitem{germandata1}
Robert Koch-Institut.
\newblock Covid-19 reports, 2020.
\newblock available at
  \url{https://www.rki.de/EN/Content/infections/epidemiology/outbreaks/COVID-19/COVID19.html}.

\bibitem{germandata2}
Center for Disaster~Management Risklayer~GmbH and Risk Reduction~Technology
  (CEDIM).
\newblock Covid-19 case numbers for germany, 2020.
\newblock available at \url{https://github.com/jgehrcke/covid-19-germany-gae}.

\bibitem{UKgov}
Gov.UK.
\newblock Covid-19 in the uk, 2020.
\newblock available at \url{https://coronavirus.data.gov.uk/details/download}.

\bibitem{CovidTracking}
CovidTracking.
\newblock The covid tracking project, 2020.
\newblock available at \url{https://covidtracking.com/data/api}.

\bibitem{havers2020seroprevalence}
Fiona~P Havers, Carrie Reed, Travis Lim, Joel~M Montgomery, John~D Klena,
  Aron~J Hall, Alicia~M Fry, Deborah~L Cannon, Cheng-Feng Chiang, Aridth
  Gibbons, et~al.
\newblock Seroprevalence of antibodies to sars-cov-2 in 10 sites in the united
  states, march 23-may 12, 2020.
\newblock {\em JAMA internal medicine}, 180(12):1576--1586, 2020.

\bibitem{bajema2020comparison}
Kristina~L Bajema, F~Scott Dahlgren, Travis~W Lim, Nicolette Bestul, Holly~M
  Biggs, Jacqueline~E Tate, Claudio Owusu, Christine~M Szablewski, Cherie
  Drenzek, Jan Drobeniuc, et~al.
\newblock Comparison of estimated sars-cov-2 seroprevalence through commercial
  laboratory residual sera testing and a community survey.
\newblock {\em Clinical Infectious Diseases}, 2020.

\bibitem{dashboard}
Centers for Disease~Control and Prevention.
\newblock Commercial laboratory seroprevalence survey data.
\newblock available at
  \url{https://covid.cdc.gov/covid-data-tracker/#national-lab}, 2020-01-04.

\bibitem{le2020sars}
Nina Le~Bert, Anthony~T Tan, Kamini Kunasegaran, Christine~YL Tham, Morteza
  Hafezi, Adeline Chia, Melissa Hui~Yen Chng, Meiyin Lin, Nicole Tan, Martin
  Linster, et~al.
\newblock Sars-cov-2-specific t cell immunity in cases of covid-19 and sars,
  and uninfected controls.
\newblock {\em Nature}, 584(7821):457--462, 2020.

\bibitem{ng2020pre}
Kevin Ng, Nikhil Faulkner, Georgina Cornish, Annachiara Rosa, Christopher Earl,
  Antoni Wrobel, Donald Benton, Chloe Roustan, William Bolland, Rachael
  Thompson, et~al.
\newblock Pre-existing and de novo humoral immunity to sars-cov-2 in humans.
\newblock {\em BioRxiv}, 2020.

\bibitem{lipsitch2020cross}
Marc Lipsitch, Yonatan~H Grad, Alessandro Sette, and Shane Crotty.
\newblock Cross-reactive memory t cells and herd immunity to sars-cov-2.
\newblock {\em Nature Reviews Immunology}, 20(11):709--713, 2020.

\bibitem{Ng2020}
Kevin~W. Ng, Nikhil Faulkner, Georgina~H. Cornish, Annachiara Rosa, Ruth
  Harvey, Saira Hussain, Rachel Ulferts, Christopher Earl, Antoni~G. Wrobel,
  Donald~J. Benton, Chloe Roustan, William Bolland, Rachael Thompson, Ana
  Agua-Doce, Philip Hobson, Judith Heaney, Hannah Rickman, Stavroula
  Paraskevopoulou, Catherine~F. Houlihan, Kirsty Thomson, Emilie Sanchez,
  Gee~Yen Shin, Moira~J. Spyer, Dhira Joshi, Nicola O~Reilly, Philip~A. Walker,
  Svend Kjaer, Andrew Riddell, Catherine Moore, Bethany~R. Jebson, Meredyth
  Wilkinson, Lucy~R. Marshall, Elizabeth~C. Rosser, Anna Radziszewska, Hannah
  Peckham, Coziana Ciurtin, Lucy~R. Wedderburn, Rupert Beale, Charles Swanton,
  Sonia Gandhi, Brigitta Stockinger, John McCauley, Steve~J. Gamblin, Laura~E.
  McCoy, Peter Cherepanov, Eleni Nastouli, and George Kassiotis.
\newblock Preexisting and de novo humoral immunity to sars-cov-2 in humans.
\newblock {\em Science}, 2020.

\bibitem{doshi2020covid}
Peter Doshi.
\newblock Covid-19: Do many people have pre-existing immunity?
\newblock {\em Bmj}, 370, 2020.

\bibitem{organisation2020territorial}
Organisation for Economic Co-operation and Development (OECD).
\newblock The territorial impact of covid-19: managing the crisis across levels
  of government, 2020.

\bibitem{gans2020economic}
Joshua~S Gans.
\newblock The economic consequences of r= 1: Towards a workable behavioural
  epidemiological model of pandemics.
\newblock {\em National Bureau of Economic Research Working Paper Series},
  2020.

\bibitem{crocerossa}
Linda~Laura Sabbadini, Maria~Clelia Romano, and Orietta Luzi.
\newblock Istat (august 03,2020) primi risultati indagine di sieroprevalenza
  sul sarscov2, 2020.
\newblock retrieved from \url{https://www.istat.it/it/archivio/246156}.

\end{thebibliography}
	\addcontentsline{toc}{chapter}{Bibliography}	 
	
\end{document}